\documentclass{wscpaperproc}
\usepackage{latexsym}
\usepackage{graphicx}
\usepackage{mathptmx}
\usepackage[T1]{fontenc}
\usepackage{pgfplots}

\usepackage{amsmath}
\usepackage{amsfonts}
\usepackage{amssymb}
\usepackage{amsbsy}
\usepackage{amsthm}
\usepackage{tikz}
\usetikzlibrary{positioning, arrows.meta, decorations.markings, calc, intersections}
\usepackage[skip=0pt]{caption}
\usepackage{subcaption}
\usepackage{diagbox}
\def\R{{\mathbb{R}}}   %set of numbers

\def\N{{\mathbb{N}}}
\def\E{{\mathbb{E}}}

\def\1{{\mathbf{1}}}

\DeclareMathOperator*{\diver}{div}
\DeclareMathOperator*{\tr}{trace}

\usepackage[pdftex,colorlinks=true,urlcolor=blue,citecolor=black,anchorcolor=black,linkcolor=black]{hyperref}

\newtheoremstyle{wsc}% hnamei
{3pt}% hSpace abovei
{3pt}% hSpace belowi
{}% hBody fonti
{}% hIndent amounti1
{\bf}% hTheorem head fontbf
{}% hPunctuation after theorem headi
{.5em}% hSpace after theorem headi2
{}% hTheorem head spec (can be left empty, meaning `normal')i

\theoremstyle{wsc}
\newtheorem{theorem}{Theorem}

\newtheorem{proposition}{Proposition}

\newtheorem{definition}{Definition}

\newtheorem{lemma}{Lemma}

\usepackage{cleveref}

\crefname{assumption}{assumption}{assumptions}

\begin{document}

%***************************************************************************
% AUTHOR: AUTHOR NAMES GO HERE
% FORMAT AUTHORS NAMES Like: Author1, Author2 and Author3 (last names)
%
%		You need to change the author listing below!
%               Please list ALL authors using last name only, separate by a comma except
%               for the last author, separate with "and"
%

% setting up general page style
\pagestyle{fancyplain}

% setting up page style of first page
\thispagestyle{plain}
\firstPageHead{}

% setting up running header (authors) of subsequent pages
\chead{\fancyplain{}{\itshape Ren and Fu}}

% setting up seperation parameters
%\headsep=72pt
\rhead{}
\cfoot{}
\renewcommand{\headrulewidth}{0pt} % (renewcommand needed in fancyhdr to remove top decorative line)
%\headrulewidth=0pt  % ("setlength" needed in fancyheading to remove top decorative line)

%%%%%%%%%%%%%%%%%%%%%%%%%%%%%%%%%%%%%%%%%%%%%%%%%%%%%%%%%%%%%%%%%%%%%%%%%%%%%%
%                                                                            %
%     THESE COMMANDS ARE REQUIRED TO WORK WITH WSC.BST TO MAKE BIBLIO     %
%                                                                            %
%%%%%%%%%%%%%%%%%%%%%%%%%%%%%%%%%%%%%%%%%%%%%%%%%%%%%%%%%%%%%%%%%%%%%%%%%%%%%%
\makeatletter
\let\@internalcite\cite
\def\cite{\def\@citeseppen{-1000}%
    \def\@cite##1##2{(##1\if@tempswa , ##2\fi)}%
    \def\citeauthoryear##1##2##3{##1 ##3}\@internalcite}
\def\citeNP{\def\@citeseppen{-1000}%
    \def\@cite##1##2{##1\if@tempswa , ##2\fi}%
    \def\citeauthoryear##1##2##3{##1 ##3}\@internalcite}
\def\citeN{\def\@citeseppen{-1000}%
%  Pierre L'Ecuyer's fix for multiple cite bug
%  Added by Paul J Sanchez on 4 October 2001
%   \def\@cite##1##2{##1\if@tempswa , ##2)\else{)}\fi}%
%   \def\citeauthoryear##1##2##3{##1 (##3}\@citedata}
    \def\@cite##1##2{##1\if@tempswa, ##2)\else{}\fi}%
    \def\citeauthoryear##1##2##3{##1 (##3)}\@citedata}
\def\citeA{\def\@citeseppen{-1000}%
    \def\@cite##1##2{(##1\if@tempswa , ##2\fi)}%
    \def\citeauthoryear##1##2##3{##1}\@internalcite}
\def\citeANP{\def\@citeseppen{-1000}%
    \def\@cite##1##2{##1\if@tempswa , ##2\fi}%
    \def\citeauthoryear##1##2##3{##1}\@internalcite}
\def\shortcite{\def\@citeseppen{-1000}%
    \def\@cite##1##2{(##1\if@tempswa , ##2\fi)}%
    \def\citeauthoryear##1##2##3{##2 ##3}\@internalcite}
\def\shortciteNP{\def\@citeseppen{-1000}%
    \def\@cite##1##2{##1\if@tempswa , ##2\fi}%
    \def\citeauthoryear##1##2##3{##2 ##3}\@internalcite}
\def\shortciteN{\def\@citeseppen{-1000}%
%  Pierre L'Ecuyer's fix for multiple cite bug
%  Added by Paul J Sanchez on 2 September 2002
%  should have caught this last year...
%   \def\@cite##1##2{##1\if@tempswa , ##2)\else{)}\fi}%
%   \def\citeauthoryear##1##2##3{##2 (##3}\@citedata}
% Shane G. Henderson fix for extra right bracket at end of optional material June 8, 2005
%    \def\@cite##1##2{##1\if@tempswa, ##2)\else{}\fi}%
    \def\@cite##1##2{##1\if@tempswa, ##2\else{}\fi}%
    \def\citeauthoryear##1##2##3{##2 (##3)}\@citedata}
\def\shortciteA{\def\@citeseppen{-1000}%
    \def\@cite##1##2{(##1\if@tempswa , ##2\fi)}%
    \def\citeauthoryear##1##2##3{##2}\@internalcite}
\def\shortciteANP{\def\@citeseppen{-1000}%
    \def\@cite##1##2{##1\if@tempswa , ##2\fi}%
    \def\citeauthoryear##1##2##3{##2}\@internalcite}
\def\citeyear{\def\@citeseppen{-1000}%
    \def\@cite##1##2{(##1\if@tempswa , ##2\fi)}%
    \def\citeauthoryear##1##2##3{##3}\@citedata}
\def\citeyearNP{\def\@citeseppen{-1000}%
    \def\@cite##1##2{##1\if@tempswa , ##2\fi}%
    \def\citeauthoryear##1##2##3{##3}\@citedata}
%
% \@citedata and \@citedatax:
%
% Place commas in-between citations in the same \citeyear, \citeyearNP,
% \citeN, or \shortciteN command.
% Use something like \citeN{ref1,ref2,ref3} and \citeN{ref4} for a list.
%
\def\@citedata{%
    \@ifnextchar [{\@tempswatrue\@citedatax}%
                  {\@tempswafalse\@citedatax[]}%
}

\def\@citedatax[#1]#2{%
\if@filesw\immediate\write\@auxout{\string\citation{#2}}\fi%
  \def\@citea{}\@cite{\@for\@citeb:=#2\do%
    {\@citea\def\@citea{, }\@ifundefined% by Young
       {b@\@citeb}{{\bf ?}%
       \@warning{Citation `\@citeb' on page \thepage \space undefined}}%
{\csname b@\@citeb\endcsname}}}{#1}}%

% don't box citations, separate with ; and a space
% also, make the penalty between citations negative: a good place to break.
%
\def\@citex[#1]#2{%
\if@filesw\immediate\write\@auxout{\string\citation{#2}}\fi%
  \def\@citea{}\@cite{\@for\@citeb:=#2\do%
    {\@citea\def\@citea{; }\@ifundefined% by Young
       {b@\@citeb}{{\bf ?}%
       \@warning{Citation `\@citeb' on page \thepage \space undefined}}%
{\csname b@\@citeb\endcsname}}}{#1}}%

% (from apalike.sty)
% No labels in the bibliography.
%
\def\@biblabel#1{}
\makeatother

%\newlength{\bibhang}
%\setlength{\bibhang}{2em}

% Indent second and subsequent lines of bibliographic entries. Taken
% from openbib.sty: \newblock is set to {}.
% \renewcommand{\refname}{REFERENCES}

\newdimen\bibindent
\bibindent=0.0em
% SEC: was \def\thebibliography#1{\section*{\refname\@mkboth
% SEC: was   {\uppercase{\refname}}{\uppercase{\refname}}}\list
\def\thebibliography#1{\section*{\refname}\list
   {}{\settowidth\labelwidth{[#1]}
   \leftmargin\parindent
   \itemindent -\parindent
   \listparindent \itemindent
   \itemsep 0pt
   \parsep 0pt}
   \def\newblock{}
   \sloppy
   \sfcode`\.=1000\relax}

           % Set up BiBTeX macros

% needed to make the tex document look more like the word counterpart :-(
\setlength{\baselineskip}{12.7pt}

% AUTHOR: Enter the title, all letters in upper case
\title{Generalizing the Generalized Likelihood Ratio Method through a Push-Out Leibniz Integration Approach}

% AUTHOR: Enter the authors of the article, see end of the example document for further examples
\author{\begin{center}Xingyu Ren\textsuperscript{1} and Michael C. Fu\textsuperscript{1,2}\\
[11pt]
\textsuperscript{1}Dept.~of Electrical and Computer Eng. \& Institute for System Research, University of Maryland, College Park, MD, USA\\
\textsuperscript{2}Robert H. Smith School of Business, University of Maryland, College Park, MD, USA\end{center}
}
\maketitle

\vspace{-12pt}

\section*{ABSTRACT}
We generalize the generalized likelihood ratio (GLR) method through a novel push-out Leibniz integration approach. Extending the conventional push-out likelihood ratio (LR) method, our approach allows the sample space to be parameter-dependent after the change of variables. Specifically, leveraging the Leibniz integral rule enables differentiation of the parameter-dependent sample space, resulting in a surface integral in addition to the usual LR estimator, which may necessitate additional simulation. Furthermore, our approach extends to cases where the change of variables only ``locally" exists. Notably, the derived estimator includes existing GLR estimators as special cases and is applicable to a broader class of discontinuous sample performances. Moreover, the derivation is streamlined and more straightforward, and the requisite regularity conditions are easier to understand and verify.

%We explore stochastic gradient estimation using the push-out likelihood ratio (LR) method, a technique that moves parameters of interest from the performance measure to the density function via a change of variables. We specifically examine scenarios where the support of the new random variable depends on the parameter, utilizing the Leibniz integral rule. In addition to the conventional LR estimator, a surface integral term is introduced to estimate the sensitivity concerning the parameter-dependent support, which may necessitate additional simulation. Furthermore, we extends these results to cases where the change of variables only exists ``locally".

\section{INTRODUCTION}
\label{sec:intro}
Consider an output sample performance parameterized by a real-valued scalar $\theta\in\Theta$: 
\begin{align*}
    \psi(X,\theta),
\end{align*}
where $\Theta$ is an open interval, $\psi:\R\times\Theta\mapsto\R$ is a real-valued function, and $X$ is the input random variable with density $f(x,\theta)$ and support $\Omega\subset\R$ (independent of $\theta$). Suppose that we are interested in estimating the derivative of the expected sample performance with respect to (w.r.t.) $\theta$:
\begin{align*} 
    \E(\psi(X,\theta))=\int_\Omega \psi(x,\theta) f(x,\theta) dx.
\end{align*}
Typical methods include infinitesimal perturbation analysis (IPA), smoothed perturbation analysis (SPA), the likelihood ratio (LR) method, and weak derivatives (WD) \cite{fu2012conditional,glasserman1990gradient,glynn1987likelilood,pflug2012optimization}. Assume that $\psi$ and $f$ are differentiable w.r.t. $\theta$, and density $f$ is absolutely continuous w.r.t. a density $f_0:\Omega\mapsto\R$ independent of $\theta$. Under suitable conditions, we can interchange the order of differentiation and integration:
\begin{align*}
    \frac{d}{d\theta} \E(\psi(X,\theta))=\int_\Omega \frac{d}{d\theta}\left(\psi(x,\theta) \frac{f(x,\theta)}{f_0(x)}\right) f_0(x) dx=\int_\Omega \left(\partial_\theta\psi(x,\theta) h(x,\theta) + \psi(x,\theta)\partial_\theta h(x,\theta) \right) f_0(x) dx,
\end{align*}
where $h(x,\theta):=f(x,\theta)/f_0(x)$ is the Radon-Nikodym derivative of $f$ w.r.t. $f_0$. With $X$ sampled from density $f_0$, $\partial_\theta\psi(X,\theta) h(X,\theta) + \psi(X,\theta)\partial_\theta h(X,\theta)$ is an example of the IPA-LR estimator \cite{l1990unified}, where $\partial_\theta\psi(X,\theta) h(X,\theta)$ and $\psi(X,\theta)\partial_\theta h(X,\theta)$ are IPA and LR estimators, respectively.

In some practical scenarios, $\psi$ is not continuous w.r.t. $\theta$ (e.g., an indicator function), or not analytically available. Consequently, differentiation cannot be passed through integration, or the partial derivative of $\psi$ may not even exist. Nevertheless, in some cases, through a change of variables, we can ``push'' the parameter $\theta$ out of the function $\psi$, to circumvent the need to differentiate a discontinuous function \shortcite{rubinstein1992sensitivity,wang2012new}. Specifically, assume that there exists a real-valued function $g(x,\theta)$ which is invertible w.r.t. $x$ for each $\theta$ and differentiable w.r.t. both arguments, such that we can express $\psi(x,\theta)= \varphi (g(x,\theta))$ for some $\varphi:\R\mapsto\R$. Define a new random variable $Y=g(X,\theta)$, whose density is given by $\tilde f(y,\theta)= f(g^{-1}(y,\theta),\theta)\left| \partial_y g^{-1}(y,\theta) \right|$ supported on $\tilde \Omega\subset\R$. Make the change of variables:
\begin{align*}
    \E(\psi(X,\theta))=\int_{\tilde\Omega}  \varphi(y) \tilde f(y,\theta) dy = \E(\varphi(Y)),
\end{align*}
and the LR method applies. \shortciteN{peng2018new,peng2020generalized} extend the push-out LR method to scenarios where $g$ is only locally invertible (i.e., its Jacobian matrix $J_g$ is invertible).

Note that the push-out LR method typically requires the support $\tilde \Omega$ of $Y$ to be independent of $\theta$. Consider a toy example $\psi(X,\theta)=\1\{X<\theta\}$, where $X$ follows an exponential distribution with parameter $\theta>0$, having density $f_X(x,\theta)=\theta e^{-\theta x}$ over the support $\Omega=[0,\infty)$. The expected sample performance can be expressed as:
\begin{align*}
    \E(\1\{X<\theta\})=\int_0^\infty \1\{x<\theta\}\theta e^{-\theta x}dx. 
\end{align*}
To apply the push-out LR method, we set $Y=\frac{X}{\theta}$, which follows an exponential distribution with parameter $\theta^2$, with density $f_Y(y,\theta)=\theta^2 e^{-\theta^2 y}$ over the support $\tilde \Omega=[0,\infty)$. Make the change of variables:
\begin{align*}
    \E(\1\{X<\theta\})=\int_0^\infty \1\{x<\theta\}\theta e^{-\theta x}dx =\int_0^\infty \1\{y<1\}\theta^2 e^{-\theta^2 y}dy= \E(\1\{Y<1\}).
\end{align*}
Since both the new sample performance $ \varphi(y)=\1\{y<1\}$ and the support of $Y$ are independent of $\theta$, we can apply the LR method w.r.t. $Y$:
\begin{align*}
    \frac{d}{d\theta}\E(\1\{X<\theta\}) = \frac{d}{d\theta}\E(\1\{Y<1\}) = \E(\1\{Y<1\}\partial_\theta\log f_Y(Y,\theta)),
\end{align*}
where $\1\{Y<1\}\partial_\theta\log f_Y(Y,\theta)=\1\{Y<1\}(2/\theta-2\theta Y)$ is an unbiased derivative estimator.

Instead of setting $Y=\frac{X}{\theta}$, an alternative approach to remove $\theta$ from the indicator function is to set $Z=X-\theta$. This creates a shifted exponential random variable with density function $f_Z(z,\theta)=\theta e^{-\theta(z+\theta)}$ over the support $[-\theta,\infty)$. Due to the dependence of the support on $\theta$, the LR method cannot be directly applied. However, if we write the integral as
\begin{align*}
    \E(\1\{X<\theta\})=\int_{-\theta}^\infty \1\{z<0\} \theta e^{-\theta (z+\theta)} dz,
\end{align*}
we can apply the Leibniz integral rule to differentiate both the lower limit and the integrand simultaneously:
\begin{align*}
    &~~~~\frac{d}{d\theta}\E(\1\{X<\theta\})=\frac{d}{d\theta}\int_{-\theta}^\infty \1\{z<0\} \theta e^{-\theta (z+\theta)} dz\\
    &=\int_{-\theta}^\infty \1\{z<0\} \frac{d}{d\theta}\theta e^{-\theta (z+\theta) } dz - \1\{z<0\} \theta e^{-\theta (z+\theta) }\big|_{z=-\theta}\frac{d}{d\theta}(-\theta)
    =\E(\1\{Z<0\}\partial_\theta\log f_Z(Z,\theta)) + \theta,
\end{align*}
which leads to a standard LR estimator augmented by an extra constant term $\theta$ arising from the differentiation w.r.t. the lower limit.

%The application of the Leibniz integral rule in the above example appears to be somewhat excessive, since a simple alteration in the change of variables could render the Leibniz integral rule unnecessary and make the LR method work. However, this observation 

This example suggests that leveraging the Leibniz integral rule extends the applicability of the push-out LR method to broader settings where the support of the newly introduced random variable may depend on the parameter. Furthermore, despite the simplicity of this example, it falls outside the scope of the generalized LR (GLR) methods proposed by \shortciteN{peng2018new,peng2020generalized}, which require either the density function to vanish at the boundary of the support or the input random variables to follow a uniform distribution. In this paper, we will explore the integration of the push-out LR method with the Leibniz integral rule for an output sample performance of the form $\varphi(g(X,\theta))$, where $X$ is a random vector and a change of variables $Y=g(X,\theta)$ removes the parameter from $\varphi$. A similar idea is proposed under different regularity conditions by \citeN{puchhammer2022likelihood}, which focuses on density estimation. The rest of this paper is organized as follows. In \Cref{sec2:inv}, we formally define the output sample performance and introduce a general form of the Leibniz integral rule for multivariate integrals, subsequently applying it to the sample performance. Specifically, we demonstrate that the new estimator includes the existing GLR estimators as special cases. In \Cref{sec3:locinv}, we extend results in \Cref{sec2:inv} to cases where the function $g(x,\theta)$ is only locally invertible w.r.t. $x$ and the sample space is unbounded. \Cref{sec:simulation} presents simulation results on the example from \Cref{sec:intro}. \Cref{sec:conc} offers conclusions and future research directions.

\section{Integrating the Leibniz integral rule with the push-out LR method}\label{sec2:inv}
Consider an output sample performance $\varphi(g(X,\theta))$, where
\begin{itemize}
    \item $\varphi:\R^n\mapsto \R$ is a bounded measurable function.
    \item $g(\cdot,\cdot):\R^n\times\Theta\mapsto \R^n$ is twice continuously differentiable w.r.t. both arguments. For each $\theta$, $g(x,\theta)$ is an invertible function of $x$. $\Theta\subset\R$ is a bounded open interval.
    \item $X$ is an $n-$dimensional random vector with bounded support $\Omega\subset\R^n$ (the boundedness condition is relaxed in \Cref{sec3:locinv}).
    \item $X$ has a density function $f(\cdot,\cdot):\Omega\times\Theta\mapsto\R$, continuously differentiable w.r.t. both arguments.
\end{itemize}
Making the change of variables $y=g(x,\theta)$, we can write 
\begin{align}\label{eq:obj}
    \E(\varphi(g(X,\theta)))=\int_{g(\Omega,\theta)} \varphi(y) f(g^{-1}(y,\theta),\theta) |\det(J_{g^{-1}}(y,\theta))| dy,
\end{align}
where $g(\Omega,\theta)$ is the image of $\Omega$ under map $g$, and $J_{g^{-1}}(y,\theta)$ is the Jacobian matrix of $g^{-1}$ w.r.t. $y$, i.e., $\{J_{g^{-1}}(y,\theta)\}_{ij}=\partial_{y_j}g_i^{-1}(y,\theta)$. Both the integrand and the domain of integration in \Cref{eq:obj} involve the parameter $\theta$. The following result introduces the Leibniz integral rule that enables differentiation of the domain w.r.t. $\theta$. \Cref{thm:leibniz} is a special case of the Leibniz integral rule proved in Section 7 and 8 of \citeN{flanders1973differentiation}, and a more general version is available in \shortciteN{amann2005analysis}.
\begin{theorem}\label{thm:leibniz}
    Let $D_\theta\subset\R^n$ be a compact set. Suppose that there exists a function $\phi(\cdot,\cdot):U\times\Theta\mapsto\R^n$, where $U\subset \R^n$ is a fixed domain, such that $D_\theta=\phi(U,\theta)$. Suppose $\phi(\cdot,\cdot):\R^n\times\Theta\mapsto\R^n$ is twice continuously differentiable in both arguments, and for each $\theta$, $\phi(x,\theta)$ is an invertible function of $x$. Then, for any function $f(\cdot,\cdot):\R^n\times\Theta\mapsto \R$ continuously differentiable in both arguments,
    \begin{align*}
        \frac{d}{d\theta}\int_{D_\theta} f(x,\theta)dx = \int_{D_\theta} \left( \partial_\theta f(x,\theta) + \text{div}(f(x,\theta)\vec v(x)) \right) dx,
    \end{align*}
    where $\text{div}$ is the divergence operator, i.e., $\text{div}(F)=\sum_{i=1}^n \partial_{x_i}F_i,~F:\R^n\mapsto\R^n$, and $\vec v(x) = \partial_\theta \phi(u,\theta)|_{u=\phi^{-1}(x,\theta)}$. In particular, by the divergence theorem \cite{zorich2004mathematical}, we can write
    \begin{align*}
        \int_{D_\theta} \text{div}(f(x,\theta)\vec v(x)) dx = \int_{\partial D_\theta} f(x,\theta) \vec v(x)^T \vec n(x)  ds,
    \end{align*}
    where $\partial D_\theta$ is the boundary of $D_\theta$, $\vec n(x)$ is the outward normal vector on surface $\partial D_\theta$, and $ds$ is the area element.
\end{theorem}

The Leibniz integral rule in $\R^n$ is more intricate than in $\R$, as the boundary of the integral domain is an $(n-1)$-dimensional "moving" surface, instead of endpoints of an interval. In fluid mechanics, the Leibniz integral rule is also known as the transport theorem \cite{frankel2011geometry}. We provide a ``physical'' interpretation of the Leibniz integral rule in $\R^2$. 

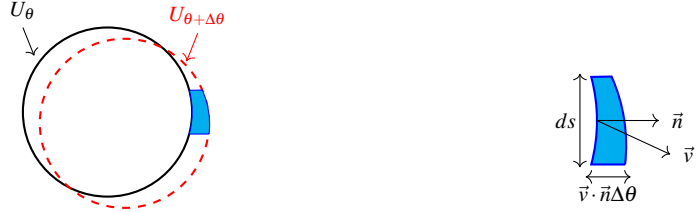
\begin{figure}[t]
    \centering
    \begin{subfigure}[b]{0.4\textwidth}
        \centering
        \begin{tikzpicture}[scale=0.75, transform shape]
            \draw[thick] (0,2) circle (1.5);
            \draw[red,thick,dashed] (0.3,1.8) circle (1.5);
            \filldraw[fill=cyan, draw=blue] (1.8,1.61) -- (1.45,1.61) arc (-15:15:1.5) -- (1.68,2.39) -- (1.68,2.39) arc (24:-6.5:1.5);
            \draw [->] (-1.5,3.5) -- (-1.3,3) node at (-1.5,3.8){$U_\theta$}; 
            \draw [->,red] (1.6,3.4) -- (1.4,2.9) node at (1.6,3.7){\textcolor{red}{$U_{\theta+\Delta\theta}$}}; 
        \end{tikzpicture}
    \end{subfigure}
    \begin{subfigure}[b]{0.4\textwidth}
        \centering
        \begin{tikzpicture}[scale=0.75, transform shape]
            \filldraw[fill=cyan, draw=blue , thick] (3.2,-0.78) -- (2.6,-0.78) arc (-15:15:3) -- (2.96,0.78) -- (2.96,0.78) arc (24:-6.5:3);
            \draw [->] (2.7,0) -- (3.8,0) node[right=0.1]{$\vec n$}; 
            \draw [->] (2.7,0) -- (4,-0.6) node[right=0.1]{$\vec v$}; 
            \draw [<->] (2.5,-1) -- (3.3,-1) node at (2.9,-1.3){$\vec v\cdot \vec n \Delta \theta$}; 
            \draw [<->] (2.4,-0.83) -- (2.4,0.83) node at (2.15,0){$ds$};
        \end{tikzpicture}       
    \end{subfigure}
    \vspace*{1mm}
    \caption{The original domain $U_\theta$ and the perturbed domain $U_{\theta+\Delta\theta}$.}\label{fig:domain}
\end{figure}

Suppose $U_\theta\subset \R^2$ is a domain with a smooth boundary and $F:\R^2\mapsto \R$ is a smooth function. We are interested in computing $\frac{d}{d\theta}\int_{U_\theta} F(x,y)dxdy$. For small $\Delta\theta$, suppose the domain $U_\theta$ moves to $U_{\theta+\Delta\theta}$, as shown in \Cref{fig:domain}. As in \Cref{thm:leibniz}, we assume that $U_\theta$ is characterized by a smooth function $\phi:\R^2\times\Theta\mapsto \R^2$ and a fixed domain $U\subset\R^2$, i.e., $U_{\theta}=\phi(U,\theta)$. Consider the difference $\int_{U_{\theta+\Delta\theta}} F(x,y)dxdy - \int_{U_\theta} F(x,y)dxdy$. The integral over the intersection $U_{\theta+\Delta\theta}\cap U_\theta$ cancels out, leaving only two strips surrounding the boundary $\partial U_\theta$ contributing to the difference. We zoom in on a small segment of this strip around a point $x\in\partial U_\theta$, illustrated by the blue region in \Cref{fig:domain}. Here, $ds$ is the arc length element, $\vec n$ is the normal vector of the boundary $\partial U_\theta$ at $x$, and $\vec v$ is the velocity vector of domain w.r.t. $\theta$, given by $\partial_\theta \phi(u,\theta)|_{u=\phi^{-1}(x,\theta)}$. For sufficiently small $\Delta\theta$ and $ds$, this region is approximately a rectangle of length $ds$ and width $\vec v\cdot \vec n \Delta \theta$, the displacement of the domain along the normal vector. Therefore, the area of the blue region is $\vec v\cdot \vec n \Delta \theta ds$, and
\begin{align*}
    \frac{d}{d\theta}\int_{U_\theta} F(x,y)dxdy= \lim_{\Delta\theta\rightarrow 0}\frac{1}{\Delta\theta} \left(\int_{U_{\theta+\Delta\theta}} F(x,y)dxdy - \int_{U_\theta} F(x,y)dxdy\right) = \int_{\partial U_\theta} F(x,y) \vec v\cdot \vec n ds.
\end{align*}
Notice that \Cref{thm:leibniz} requires that the domain $D_\theta$ to be parameterized by a sufficiently smooth function $\phi(u,\theta)$ defined on a fixed set $U$. In our formulation, these correspond to the function $g$ and the sample space $\Omega$. However, \Cref{thm:leibniz} also requires the integrand to be differentiable, a condition that $\varphi$ may not satisfy. To address this, we can approximate $\varphi$ by smooth functions.
\begin{proposition}\label{prop:dense}
    Compactly supported smooth functions are dense in $L^p(\R^n),~1\leq p<\infty$ and $C(\R^n)$ (the space of continuous functions on $\R^n$).
\end{proposition}
See \shortciteN{peng2018new} and Section 8.2 in \citeN{folland1999real} for the proof and a method for constructing smooth approximations via convolution with mollifiers. As both $\Omega$ and $\Theta$ are bounded sets, the set $g(\Omega,\Theta):=\{y\in\R^n \ | \ y=g(x,\theta), \ (x,\theta)\in\Omega\times\Theta\}$ is also bounded. In our problem formulation, we can restrict $\varphi$ to this bounded set $g(\Omega,\Theta)$. Since $\varphi$ is bounded, it is integrable over $g(\Omega,\Theta)$. By \Cref{prop:dense}, there exists a sequence of smooth functions $\{\varphi_n\}_{n\in\N}$ such that $\varphi_n\rightarrow\varphi$ in $L^1$ as $n\rightarrow \infty$. Substituting $\varphi_n,~n\in\N$ into \Cref{eq:obj}, we can apply \Cref{thm:leibniz}:
\begin{align}
        \frac{d}{d\theta}\E(\varphi_n(g(X,\theta)))&=\frac{d}{d\theta}\int_{g(\Omega,\theta)} \varphi_n(y) f(g^{-1}(y,\theta),\theta) |\det(J_{g^{-1}}(y,\theta))| dy \nonumber\\
        &=\int_{g(\Omega,\theta)} \varphi_n(y) \frac{d}{d\theta} (f(g^{-1}(y,\theta),\theta) |\det(J_{g^{-1}}(y,\theta))|) dy \label{eq:term1} \\
        &+\int_{g(\Omega,\theta)} \diver(\varphi_n(y) f(g^{-1}(y,\theta),\theta) |\det(J_{g^{-1}}(y,\theta))| \vec v(y)) dy, \label{eq:term2}
\end{align}
where $\vec v(y)=\partial_\theta g(x,\theta)|_{x=g^{-1}(y,\theta)}$. Notice that for each $\theta$, $g(\cdot,\theta):\R^n\mapsto\R^n$ is a diffeomorphism, defined as follows \cite{zorich2004mathematical1}.
\begin{definition}
    A mapping $f:U\mapsto V$, where $U, V$ are open subsets of $\R^m$, is a diffeomorphism of order $p$ if $f$ is $p$-times continuously differentiable, $f$ is a bijection, and $f^{-1}:V\mapsto U$ is $p$-times continuously differentiable.
\end{definition}
The fact that $g$ is a diffeomorphism directly follows from the following inverse function theorem which is very useful for establishing \Cref{prop:cov}.
\begin{lemma}\label{thm:inverse}
    Suppose a mapping $f:G\mapsto\R^m$ of a domain $G\subset\R^m$ is such that $f$ is $p$-times continuously differentiable, $y_0=f(x_0)$ at some $x_0\in G$, and the Jacobian matrix $J_f(x_0)$ invertible. Then there exists a neighborhood $U(x_0)\subset G$ of $x_0$ and a neighborhood $V(y_0)$ of $y_0$ such that $f:U(x_0)\mapsto V(y_0)$ is a diffeomorphism of order $p$. Moreover, if $x\in U(x_0)$ and $y=f(x)\in V(y_0)$, then $J_{f^{-1}}(y)=J_{f}^{-1}(x)$.
\end{lemma}
See Section 8.6 in \citeN{zorich2004mathematical1} for a proof of \Cref{thm:inverse}. Notice that the image set $g(\Omega,\theta)$ can be complex in high-dimensional spaces, and in some cases, the function $g$ doesn't have a closed-form inverse. Specifically, in \Cref{sec3:locinv}, we study the generalized scenario where $g$ is only locally invertible, meaning there is no global change of variables $y=g(x,\theta)$. Therefore, we would like to reverse the change of variables.
\begin{proposition}\label{prop:cov}
    For $y=g(x,\theta)$, the following equations hold:
    \begin{align}
        &\frac{d}{d\theta} (f(g^{-1}(y,\theta),\theta) |\det(J_{g^{-1}}(y,\theta))|)= |\det(J_{g^{-1}}(y,\theta))|(d(x,\theta)+l(x,\theta)) f(x,\theta)\label{eq:cov1},\\
        &\diver(\varphi_n(y) f(g^{-1}(y,\theta),\theta)|\det(J_{g^{-1}}(y,\theta))| \vec v(y)) = |\det(J_{g^{-1}}(y,\theta))| \diver(\varphi_n(g(x,\theta)) f(x,\theta)s(x,\theta))\label{eq:cov2},
    \end{align}
    where $d(x,\theta)= \text{div}(-f(x,\theta)J_g^{-1}(x,\theta)\partial_\theta g(x,\theta))/f(x,\theta)$ and $l(x,\theta)=\partial_\theta \log f(x,\theta)$ are real-valued functions, and $s(x,\theta)= J_g^{-1}(x,\theta)\partial_\theta g(x,\theta) $ is an $n-$dimensional vector-valued function.
\end{proposition}

%To compute \Cref{eq:term2}, we first notice that $\partial g(\Omega,\theta)= g(\partial\Omega,\theta)$, following from the fact that $g$ is a diffeomorphism.

%\begin{theorem}
%    Suppose $M$ and $N$ are smooth manifolds with boundary and $F:M\mapsto N$ is a diffeomorphism. Then $F(\partial M)=\partial N$ , and $F$ restricts to a diffeomorphism from Int $M$ to Int $N$.
%\end{theorem}

See the \hyperlink{sec:app1}{Appendix} for the proof. To reverse the change of variables, we substitute \eqref{eq:cov1} and \eqref{eq:cov2} into \eqref{eq:term1} and \eqref{eq:term2}, respectively:
\begin{align*}
    &\int_{g(\Omega,\theta)} \varphi_n(y) \frac{d}{d\theta} (f(g^{-1}(y,\theta),\theta) |\det(J_{g^{-1}}(y,\theta))|) dy=\int_\Omega \varphi_n(g(x,\theta))(d(x,\theta)+l(x,\theta)) f(x,\theta) dx,\\
    &\int_{ g(\Omega,\theta)} \diver (\varphi_n(y) f(g^{-1}(y,\theta),\theta) |\det(J_{g^{-1}}(y,\theta))| \vec v(y)) dy=\int_\Omega \diver (\varphi_n(g(x,\theta))f(x,\theta) s(x,\theta)) dx.
\end{align*}
By the divergence theorem, $\int_\Omega \diver (\varphi_n(g(x,\theta))f(x,\theta) s(x,\theta)) dx=\int_{\partial\Omega}\varphi_n(g(x,\theta))s(x,\theta)^T \vec n(x)f(x,\theta) ds$, where $\vec n(x)$ is the outward normal vector on surface $\partial\Omega$ \cite{zorich2004mathematical1}. To summarize, we can write
\begin{align}\label{eq:glr-approx}
    \frac{d}{d\theta}\E(\varphi_n(g(X,\theta)))= \int_\Omega \varphi_n(g(x,\theta)) (d(x,\theta)+l(x,\theta)) f(x,\theta)dx+\int_{\partial\Omega}\varphi_n(g(x,\theta))s(x,\theta)^T \vec n(x)f(x,\theta) ds.
\end{align}
Under suitable conditions, $\frac{d}{d\theta}\E(\varphi_n(g(X,\theta)))$ converges to $\frac{d}{d\theta}\E(\varphi(g(X,\theta)))$ as $n\rightarrow\infty$.
\begin{theorem}\label{thm:glr1}
    If $\lim_{n\rightarrow\infty}\int_{\partial\Omega}\sup_{\theta\in\Theta}|(\varphi(g(x,\theta))-\varphi_n(g(x,\theta)))s(x,\theta)^T \vec n(x)f(x,\theta)| ds=0$, then 
    \begin{align}\label{eq:glr}
        \frac{d}{d\theta}\E(\varphi(g(X,\theta)))= \int_\Omega \varphi(g(x,\theta)) (d(x,\theta)+l(x,\theta)) f(x,\theta)dx+\int_{\partial\Omega}\varphi(g(x,\theta))s(x,\theta)^T \vec n(x)f(x,\theta) ds,
    \end{align}
    where $d(x,\theta)= \text{div}(-f(x,\theta)s(x,\theta))/f(x,\theta)$, $l(x,\theta)=\partial_\theta \log f(x,\theta)$, and $s(x,\theta)= J_g^{-1}(x,\theta)\partial_\theta g(x,\theta) $.
\end{theorem}
See the \hyperlink{sec:app2}{Appendix} for the proof. \Cref{thm:glr1} can be extended to functions $g(\cdot,\cdot):\R^m\times\Theta\mapsto \R^n$ and $\varphi:\R^m\mapsto \R$, where $m<n$, by replacing $J_g^{-1}(x,\theta)$ with an $m\times m$ invertible submatrix of it \shortcite{peng2018new}. In \Cref{eq:glr}, the first integral on the right-hand side is derived by differentiating the density of the random variable $Y=g(X,\theta)$ and reversing the change of variables. An unbiased gradient estimator for it is given by
\begin{align}
    \varphi(g(X,\theta)) (d(X,\theta)+l(X,\theta)) \label{eq:glr1}.
\end{align}
The second term in \Cref{eq:glr} is a surface integral that arises from differentiating the domain $g(\Omega,\theta)$ w.r.t. $\theta$. If the domain $g(\Omega,\theta)$ does not depend on $\theta$, the surface integral vanishes. In general, computing the surface integral is challenging unless the surface can be parameterized and the normal vector has a closed-form expression. However, for certain special forms of $\Omega$ and $\varphi$, the surface integral can be converted into a regular integral that is easier to handle.
\subsection*{Rectangle support}
Consider the case where $\Omega=[a_1,b_1]\times\cdots\times[a_n,b_n]$, a hyperrectangle in $\R^n$, with boundary given by $\partial \Omega = \cup_{i=1}^n (\Omega_{a_i}\cup\Omega_{b_i})$, a union of surfaces where
\begin{align*}
    \Omega_{a_i}: = [a_1,b_1]\times\cdots\times\{a_i\}\times\cdots\times[a_n,b_n],~\Omega_{b_i}: = [a_1,b_1]\times\cdots\times\{b_i\}\times\cdots\times[a_n,b_n].
\end{align*}
For each $i$, the normal vector $\vec n(x)$ for surfaces $\Omega_{a_i}$ and $\Omega_{b_i}$ are $-e_i$ and $e_i$, respectively, where $e_i\in\R^n$ is the unit vector with $i^{\text{th}}$ component to be one. The surface integral over each $\Omega_{a_i}$ and $\Omega_{b_i}$ reduces to a standard multivariate integral:
\begin{align*}
    &~~~~\int_{\partial\Omega}\varphi(g(x,\theta))s(x,\theta)^T \vec n(x)f(x,\theta) ds\\
    &=\sum_{i=1}^n\int_{x_j\in[a_j,b_j],j=1,\dots,n,j\neq i} \varphi(g(x,\theta))s(x,\theta)^Te_i f(x,\theta) \prod_{j=1,\dots,n,j\neq i} dx_j \Big| ^{b_i}_{x_i=a_i}\\
    &=\sum_{i=1}^n (\E(\varphi(g(X,\theta))s(X,\theta)^Te_i|X_i=b_i)f_{X_i}(b_i)-\E(\varphi(g(X,\theta))s(X,\theta)^Te_i|X_i=a_i)f_{X_i}(a_i)),
\end{align*}
where $f_{X_i}$ is the marginal density of $X_i$. An unbiased gradient estimator for the surface integral is given by
\begin{align}
    \sum_{i=1}^n (\varphi(g(X,\theta))f_{X_i}(b_i)s(X,\theta)^Te_i  \big|_{X\sim f_{X|X_i=b_i}}-\varphi(g(X,\theta))f_{X_i}(a_i)s(X,\theta)^Te_i  \big|_{X\sim f_{X|X_i=a_i}}) \label{eq:surface},
\end{align}
where $f_{X|X_i}$ is the conditional density of $X$ given $X_i$. In particular, if $\{X_i\}_{i=1,\dots,n}$ are independent, estimator \eqref{eq:surface} simplifies to
\begin{align*}
    \sum_{i=1}^n (\varphi(g(X,\theta))f_{X_i}(b_i)s(X,\theta)^Te_i\big|_{X_i=b_i}-\varphi(g(X,\theta))f_{X_i}(a_i)s(X,\theta)^Te_i\big|_{X_i=a_i}),
\end{align*}
which can be simulated by a single sample path, concurrently with estimator \eqref{eq:glr1}. \shortciteN{peng2020generalized} studies a special case where the input consists of an independent sequence of uniform random variables. Another special case occurs when the density function vanishes at the boundary of the support \shortcite{peng2018new}. For the latter case, the marginal densities $f_{X_i}(a_i)$ and $f_{X_i}(b_i)$ are zero, resulting in the surface integral vanishing, as well.

\subsection*{Almost everywhere (a.e.) differentiable $\varphi$}
\renewcommand{\thefootnote}{\fnsymbol{footnote}}
For an a.e. differentiable function $F:\R^n\mapsto\R^n$ with set of discontinuities $D_F$, the divergence theorem holds under the certain conditions \cite{shapiro1958divergence}. Suppose $\Gamma\subset\R^n$ is a bounded set and its boundary $\partial \Gamma$ is a simple closed curve. If the following conditions hold:
\begin{itemize}
    \item $F$ is continuous on closure$(\Gamma)\setminus D_F$ and is $L^2$-integrable on $\Gamma$.
    \item $\diver F$ exists a.e. and is integrable on $\Gamma$.
    \item $\diver_* F$ and $\diver^* F$ are finite on $\Gamma\setminus D_F$, with
        \begin{align*}
            \text{div} _* F(y):=\liminf_{t\rightarrow 0}\frac{1}{\text{vol}(B(y,t))}\int_{\partial B(y,t)} F(y)^T \vec n(y)dy,
        \end{align*}
    where $B(y,t)=\{y'\in\R^n \ | \ \|y'-y\|_\infty<t \}$ is an open ball centered at $y$ with radius $t$, and $\text{vol}(B(y,t))$ is its $n-$dimensional volume. $\diver^* F$ is defined similarly by replacing $\liminf$ with $\limsup$.
    \item The set $D_F$ has logarithmic capacity zero if $n=2$, or Newtonian capacity zero if $n\geq3$. For a compact set $K$, the logarithmic capacity is given by $\exp\left(-\min_\mu \int_K\int_K \log(|x-y|^{-1})d\mu(x)d\mu(y)\right)$, and the Newtonian capacity is given by $\left(\min_\mu \int_K\int_K |x-y|^{-(n-2)}d\mu(x)d\mu(y)\right)^{-1}$, where the minimum is taken over all Borel probability measures on $K$ \cite{landkof1972foundations}.
\end{itemize}
Then, the divergence theorem holds on $\Gamma$: $\int_\Gamma \diver(F(y))dy =\int_{\partial\Gamma}F(y)^T \vec n(y)dy$. Notice that the condition ``Newtonian capacity zero'' is stronger than the condition ``measure zero''. For example, in $\R^3$, both a two-dimensional disk and a line segment have Lebesgue measure zero. However, the line segment has zero Newtonian capacity, whereas the two-dimensional disk has a positive capacity \cite{landkof1972foundations}.

Suppose that $\varphi$ is bounded and differentiable a.e. except on a set of capacity zero. Since functions $g,f$ and $s$ are continuously differentiable, the divergence theorem holds:
\begin{align*}
    \int_\Omega \diver(\varphi(g(x,\theta))s(x,\theta)f(x,\theta))dx =\int_{\partial\Omega} \varphi(g(x,\theta))s(x,\theta)^T \vec n(x) f(x,\theta)dx.
\end{align*}
Clearly, $\diver(\varphi(g(X,\theta))s(X,\theta)f(X,\theta))/f(X,\theta)$ is an unbiased estimator for the surface integral. Combined with estimator \eqref{eq:glr1}, we obtain a single-run unbiased estimator for $\frac{d}{d\theta}\E(\varphi(g(X,\theta)))$:
\begin{align*}
    \varphi(g(X,\theta)) (d(X,\theta)+l(X,\theta))+\diver(\varphi(g(X,\theta))s(X,\theta)f(X,\theta))/f(X,\theta).
\end{align*}

%Consider the case where $\varphi(g(x,\theta))=\1\{g(x,\theta)>0\}$. Then,
%\begin{align*}
%    &~~~~\int_{\partial\Omega}\varphi(g(x,\theta))s(x,\theta)^T \vec n(x)f(x,\theta) ds\\
%    &
%\end{align*}

\section{Local change of variables via the inverse function theorem}\label{sec3:locinv}
In this section, we relax the condition for $g$ to be invertible everywhere and instead consider it being locally invertible. Specifically, we only assume that its Jacobian matrix $J_g$ is invertible a.e., which is a necessary but not sufficient condition for global invertibility. By \Cref{thm:inverse}, except on a set of measure zero, for each $x\in\Omega$, there exists a bounded open neighborhood $U(x)$ of $x$, such that $g(\cdot,\theta)$ is invertible on $U(x)$. Since $\Omega$ is bounded, by the Heine-Borel theorem, there exists a finite collection of open neighborhoods $\{U_i\}_{i=1,\cdots,N}$, such that closure$(\Omega)\subset \cup_{i=1}^N U_i$. For each $i$, we can derive a ``local'' version of \Cref{eq:glr-approx} over $\Omega\cap U_i$:
\begin{align*}
    &~~~~\frac{d}{d\theta}\E(\varphi_n(g(X,\theta))\1\{X\in(\Omega\cap U_i)\})\\
    &=\int_{\Omega\cap U_i} \varphi_n(g(x,\theta)) (d(x,\theta)+l(x,\theta)) f(x,\theta)
    +\diver(\varphi_n(g(x,\theta))s(x,\theta)f(x,\theta)) dx.
\end{align*}
Combining all the open sets $\{U_i\}_{i=1,\cdots,N}$, we can reconstruct \Cref{eq:glr-approx} over the entire sample space $\Omega$:
\begin{align*}
    \frac{d}{d\theta}\E(\varphi_n(g(X,\theta)))&= \sum_{i=1}^N \int_{\Omega\cap U_i} \varphi_n(g(x,\theta)) (d(x,\theta)+l(x,\theta)) f(x,\theta)+\diver(\varphi_n(g(x,\theta))s(x,\theta)f(x,\theta)) dx\\
    &=\int_{\Omega} \varphi_n(g(x,\theta)) (d(x,\theta)+l(x,\theta)) f(x,\theta)+\diver(\varphi_n(g(x,\theta))s(x,\theta)f(x,\theta)) dx\\
    &=\int_\Omega \varphi_n(g(x,\theta)) (d(x,\theta)+l(x,\theta)) f(x,\theta)dx+\int_{\partial\Omega}\varphi_n(g(x,\theta))s(x,\theta)^T \vec n(x)f(x,\theta) ds.
\end{align*}
Therefore, the proof of \Cref{thm:glr1} still holds for locally invertible function $g$.

\subsection*{Unbounded sample space $\Omega$}
In addition to the local change of variables, \Cref{thm:glr1} can be extended to the unbounded sample space $\Omega$ under appropriate conditions. We provide a brief outline of this extension, leaving the detailed exploration to future research. Consider $\Omega_L:=\Omega\cap[-L,L]^n$, the restriction of $\Omega$ to the hyperrectangle $[-L,L]^n$. For a fixed $L>0$, by \Cref{prop:dense}, there exists a sequence of smooth functions $\{\varphi_{n,L}\}_{n\in\N}$ such that $\varphi_{n,L}\rightarrow\varphi$ in $L^1$ as $n\rightarrow \infty$ over the compact set $g(\Omega_L,\Theta)$. Hence, we can reconstruct \Cref{eq:glr-approx} over $\Omega_L$:
\begin{align*}
    &~~~~\frac{d}{d\theta}\E(\varphi_{n,L}(g(X,\theta))\1\{X\in\Omega_L\}) \\
    &= \int_{\Omega_L}\varphi_{n,L}(g(x,\theta)) (d(x,\theta)+l(x,\theta)) f(x,\theta)dx+\int_{\partial\Omega_L}\varphi_{n,L}(g(x,\theta))s(x,\theta)^T \vec n(x)f(x,\theta) ds.
\end{align*}
Our goal is to show $\lim_{n \rightarrow \infty}\frac{d}{d\theta}\E(\varphi_{n,L}(g(X,\theta)))=\frac{d}{d\theta}\E(\varphi(g(X,\theta))\1\{X\in\Omega_L\})$. By \Cref{thm:glr1}, a sufficient condition is $\lim_{n\rightarrow\infty}\int_{\partial\Omega_L}\sup_{\theta\in\Theta}|(\varphi(g(x,\theta))\1\{X\in\Omega_L\}-\varphi_{n,L}(g(x,\theta)))s(x,\theta)^T \vec n(x)f(x,\theta)| ds=0$. Taking $n \rightarrow \infty$, we obtain
\begin{align}\label{eq:compact}
    \begin{split}
        &~~~~\frac{d}{d\theta}\E(\varphi(g(X,\theta))\1\{X\in\Omega_L\})\\
        &= \int_{\Omega_L}\varphi(g(x,\theta)) (d(x,\theta)+l(x,\theta)) f(x,\theta)dx+\int_{\partial\Omega_L}\varphi(g(x,\theta))s(x,\theta)^T \vec n(x)f(x,\theta) ds.    
    \end{split}
\end{align}
Next, we aim to show that $\lim_{L\rightarrow\infty}\frac{d}{d\theta}\E(\varphi(g(X,\theta))\1\{X\in\Omega_L\})=\frac{d}{d\theta}\E(\varphi(g(X,\theta)))$, for which a sufficient condition is the uniform convergence of both integrals on the right-hand side of \Cref{eq:compact} over $\Theta$ as $L\rightarrow\infty$. Specifically, a sufficient condition for the uniform convergence of the first integral is $\int_{\Omega}\sup_{\theta\in\Theta}|\varphi(g(x,\theta)) (d(x,\theta)+l(x,\theta)) |f(x,\theta)dx<\infty$. We refer to Theorem 4 in Section 16.3.5 of \citeN{zorich2004mathematical} for the conditions under which the interchange of the order of limit and integral is permissible.

%We have the option to approach the limit with either $n\rightarrow\infty$ first or $L\rightarrow\infty$ first. Consequently, this yields two distinct sets of regularity conditions for \Cref{thm:glr1} to hold. If taking $L\rightarrow\infty$ first, we want
%\begin{align*}
%    \frac{d}{d\theta}\E(\varphi_n(g(X,\theta)))= \int_{\Omega}\varphi_n(g(x,\theta)) (d(x,\theta)+l(x,\theta)) f(x,\theta)dx+\int_{\Omega}\diver(\varphi_n(g(x,\theta))s(x,\theta) f(x,\theta)) dx,
%\end{align*}
%for which a sufficient condition is that the integrand is uniformly integrable over $\Theta$. Then we need to show that $\lim_{n\rightarrow\infty}\frac{d}{d\theta}\E(\varphi_n(g(X,\theta)))\rightarrow\frac{d}{d\theta}\E(\varphi(g(X,\theta)))$.

%\begin{theorem}\label{thm:glr2}
%    If the following conditions hold:
%    \begin{align}
%        &\int_{\Omega}\sup_{\theta\in\Theta}|\varphi(g(x,\theta)) (d(x,\theta)+l(x,\theta)) |f(x,\theta)dx<\infty\nonumber,\\
%        &\lim_{L\rightarrow\infty}\sup_{\theta\in\Theta} \left|\int_{\partial\Omega_L}\varphi(g(x,\theta))s(x,\theta)^T \vec n(x)f(x,\theta) ds-\int_{\partial\Omega}\varphi(g(x,\theta))s(x,\theta)^T \vec n(x)f(x,\theta) ds\right|=0\label{}.
%    \end{align}
%    Then,
%    \begin{align*}
%        \frac{d}{d\theta}\E(\varphi(g(X,\theta)))= \int_\Omega \varphi(g(x,\theta)) (d(x,\theta)+l(x,\theta)) f(x,\theta)dx+\int_{\partial\Omega}\varphi(g(x,\theta))s(x,\theta)^T \vec n(x)f(x,\theta) ds.
%    \end{align*}
%\end{theorem}

\section{Simulation example}\label{sec:simulation}
In this section, we evaluate the generalized GLR method using the toy example introduced in \Cref{sec:intro}: $\E(\1\{X<\theta\})$, where $X$ follows an exponential distribution with parameter $\theta>0$. Notice that its derivative can be computed analytically:
\begin{align*}
    \frac{d}{d\theta}\E(\1\{X<\theta\}) =\frac{d}{d\theta} \int_0^\theta \theta e^{-\theta x}dx= (\theta e^{-\theta x})|_{x=\theta}\frac{d}{d\theta}(\theta) +\int_0^\theta \frac{d}{d\theta} (\theta e^{-\theta x}) dx=2\theta e^{-\theta^2}.
\end{align*}
Using the conventional push-out LR method, we obtain an unbiased estimator as follows:
\begin{align*}
    \frac{d}{d\theta}\E(\1\{X<\theta\})_{\text{LR}}=\E \left(\1\{Y<1\}\partial_\theta\log f_Y(Y,\theta)\right)=\E \left(\1\{Y<1\}(2/\theta-2\theta Y)\right),
\end{align*}
where $Y$ follows an exponential distribution with parameter $\theta^2$. The method introduced in \Cref{thm:glr1} is referred to as the GLR* method, offering another unbiased estimator:
\begin{align*}
    \frac{d}{d\theta}\E(\1\{X<\theta\})_{\text{GLR*}}=\E\left(\1\{X<\theta
    \}(d(X,\theta)+l(X,\theta))+\theta\right)=\E\left(\1\{X<\theta
    \}(1/\theta-\theta-X)+\theta\right),
\end{align*}
where the constant $\theta$ corresponds to estimator \eqref{eq:surface}, the derivative of the parameter-dependent domain. 

\begin{figure}[h]
    \begin{minipage}{.4\textwidth}
    \subfloat{
    \begin{tikzpicture}[scale=0.75, transform shape]
        \begin{axis}[
            ybar,
            bar width=7pt,
            xlabel={$\theta$},
            %ylabel={$\frac{d}{d\theta}\E(\1\{X<\theta\})$},
            %ylabel near ticks,
            %ylabel style={rotate=-90},
            % (changed to an absolute value)
            enlarge x limits={abs=0.1},
            % ---------------------------------------------------------------------
            % changes to get what you want
            % ---------------------------------------------------------------------
            ymin=0.3,
            scaled ticks=false,
            % remove the `xticks`
            %xtick style={
            %    /pgfplots/major tick length=0pt,
            %},
            xtick=data,
            %xtick distance=0.25,
            legend style={at={(0.23,0.96)},
		    anchor=north,legend columns=1},
            legend image code/.code={
            \draw [#1] (0cm,-0.1cm) rectangle (0.2cm,0.25cm); },
        ]
        \addplot 
            coordinates {(0.2,0.3843) (0.4,0.6817)
                 (0.6,0.8372) (0.8,0.8437) };
        \addplot+ [
            error bars/.cd,
            y dir=both,
            % (changed from `y explicit` so the error bars are (clearly) visible
            y explicit relative,
            ] 
            coordinates {
                    (0.2,0.3878) +- (0,0.0184)
                    (0.4,0.6825) +- (0,0.0136)
                    (0.6,0.8488) +- (0,0.0075)
                    (0.8,0.8418) +- (0,0.0033)};
        \addplot+ [
            error bars/.cd,
            y dir=both,
            % (changed from `y explicit` so the error bars are (clearly) visible
            y explicit relative,
            ] 
            coordinates {
                    (0.2,0.3769) +- (0,0.0378)
                    (0.4,0.6637) +- (0,0.0324)
                    (0.6,0.8495) +- (0,0.0259)
                    (0.8,0.8530) +- (0,0.0188)};
        \legend{True value,GLR*,Push-Out LR}
        \end{axis}
        \node[above] at (0.3,5.9) {$\frac{d}{d\theta}\E(\1\{X<\theta\})$};
    \end{tikzpicture}
    }
    \end{minipage}
    \begin{minipage}{.4\textwidth}
    \subfloat{
        \begin{tabular}{cccc}
            \hline\hline
            $\theta$ &True value &GLR* &Push-Out LR \\ \hline\hline
            $0.2$ &$0.384$ &$0.388\pm0.018$ &$0.377\pm0.038$ \\ \hline
            $0.4$ &$0.682$ &$0.683\pm0.014$ &$0.664\pm0.032$ \\ \hline
            $0.6$ &$0.837$ &$0.849\pm0.008$ &$0.850\pm0.026$ \\ \hline
            $0.8$ &$0.844$ &$0.842\pm0.003$ &$0.853\pm0.019$ \\ \hline
        \end{tabular}
    }
    \end{minipage}
    \caption{Simulation results: Point estimates and standard errors for $\frac{d}{d\theta}\E(\1\{X<\theta\})$.}\label{fig:simulation}
\end{figure}
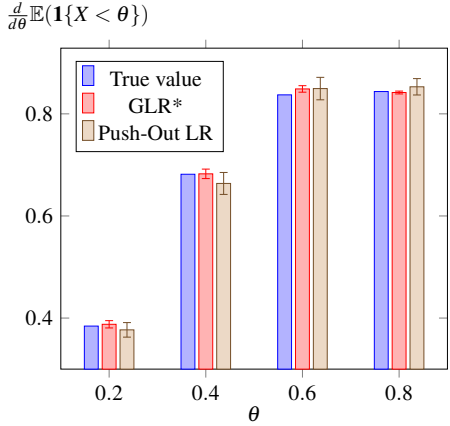

We simulate both derivative estimators at $\theta=0.2,0.4,0.6,0.8$ with $2500$ independent replications. The simulation results are depicted in \Cref{fig:simulation}. Both estimators demonstrate satisfactory accuracy. Notably, the standard errors of the GLR* estimator are half or even less of those of the push-out LR estimator. This observation can be explained as follows. From \Cref{eq:glr}, we observe that the ``randomness'' of the derivative is split into two components. One component represents a conventional LR estimator (after the change of variables). The other component captures the sensitivity the integration domain w.r.t. $\theta$, and \eqref{eq:surface} is an unbiased estimator for this component. In this simple example, the latter component is merely the constant term $\theta$, whereas if the input $X$ is a random vector, additional simulations might be required to estimate the value of the surface integral. Thus, the reduction in variance comes at the expense of potentially additional simulation runs. Moreover, for this example, it is more efficient to use the conditional density estimator \shortcite{l2022monte}.

\section{Conclusion}\label{sec:conc}
In this paper, we introduce a novel push-out Leibniz integration approach to generalize the GLR method. The underlying idea of our method is straightforward: ``push'' the parameter $\theta$ out of the performance measure $\varphi(g(X,\theta))$ through a change of variables $Y=g(X,\theta)$, differentiate the transformed density function $f_Y$ and integration domain $g(\Omega,\theta)$ using the Leibniz integral rule, and finally reverse the change of variables $X=g^{-1}(Y,\theta)$. Compared to the push-out LR method, the newly derived estimator can be applied to a wider range of gradient estimation problems where the sample space is parameter-dependent and the function $g$ is only locally invertible. We demonstrate that the newly derived estimator encompasses the existing GLR estimators as special cases. Simulation results suggest that the generalized GLR estimator, compared to the push-out LR method, can reduce variance at the expense of potentially additional simulations. For future research, we aim to extend our results from compact sample spaces to unbounded sample spaces and apply them to more practical scenarios. We also observe that the form of the estimator \eqref{eq:surface} for the surface integral resembles the form of some SPA estimators \cite{fu2012conditional}. Investigating the connection between GLR and SPA estimators is an interesting direction for further research.

\section*{ACKNOWLEDGMENTS}
This work was supported in part by the National Science Foundation under Grant IIS-2123684 and by AFOSR under Grant FA95502010211.

\appendix

\section*{APPENDIX}
\subsection*{Proof of \Cref{prop:cov}} 
We refer to Chapter 2 from \citeN{frankel2011geometry} for the justification of \Cref{eq:diver,eq:diver1}.\\
\textbf{\Cref{eq:cov1}:} \hypertarget{sec:app1}{By} the chain rule,
\begin{align}
    \begin{split}\label{eq:cov1_1}
        \frac{d}{d\theta} (f(g^{-1}(y,\theta),\theta) |\det(J_{g^{-1}}(y,\theta))|)&= (\nabla_x f(g^{-1}(y,\theta),\theta)^T \partial_\theta g^{-1}(y,\theta)+ \partial_\theta f(g^{-1}(y,\theta),\theta))\\
        &\times |\det(J_{g^{-1}}(y,\theta))| +f(g^{-1}(y,\theta),\theta)\partial_\theta |\det(J_{g^{-1}}(y,\theta))|,    
    \end{split}
\end{align}
Notice that $g(g^{-1}(y,\theta),\theta)=y$. Therefore, by (implicit) differentiation,
\begin{align*}
    0=\frac{d}{d\theta}g(g^{-1}(y,\theta),\theta)=\partial_\theta g(g^{-1}(y,\theta),\theta) + J_g(g^{-1}(y,\theta),\theta)\partial_\theta g^{-1}(y,\theta),
\end{align*}
i.e.,
\begin{align} \label{eq:t1}
    \partial_\theta g^{-1}(y,\theta) = -J_g^{-1}(g^{-1}(y,\theta),\theta) \partial_\theta g(g^{-1}(y,\theta),\theta).
\end{align}
To compute $\partial_\theta |\det(J_{g^{-1}}(y,\theta))|$, we use the fact that $\partial_\theta\det A(\theta)=\det A(\theta) \text{trace}(\partial_\theta A(\theta) A(\theta)^{-1} )$. Since $g$ is twice continuously differentiable, $\partial_\theta\partial_{ y_j} g_i^{-1}(y,\theta)=\partial_{y_j}(\partial_\theta g_i^{-1}(y,\theta))$, and
\begin{align*}
    \text{trace}\left( (\partial_\theta J_{g^{-1}}(y,\theta)) J_{g^{-1}}(y,\theta)^{-1}\right) = \sum_{i,j}^n \partial_{y_j}(\partial_\theta g_i^{-1}(y,\theta)) \partial_{x_i} g_j(x,\theta)|_{x=g^{-1}(y,\theta)} = \sum_{i=1}^n \partial_{x_i}(\partial_\theta g_i^{-1}(y,\theta))|_{y=g(x,\theta)}.
\end{align*}
It follows that
\begin{align} \label{eq:t2}
    \partial_\theta \det(J_{g^{-1}}(y,\theta)) = \det(J_{g^{-1}}(y,\theta)) \left(\sum_{i=1}^n \partial_{x_i}(\partial_\theta g_i^{-1}(y,\theta))|_{y=g(x,\theta)}\right).
\end{align}
Substituting \Cref{eq:t1,eq:t2} into \Cref{eq:cov1_1}, we obtain \Cref{eq:cov1}:
\begin{align*}
    &~~~~\frac{d}{d\theta} (f(g^{-1}(y,\theta),\theta) |\det(J_{g^{-1}}(y,\theta))|) dy\\
    &= (-\nabla_x f(x,\theta)^T J_g^{-1}(x,\theta) \partial_\theta g(x,\theta)+ \partial_\theta f(x,\theta) )|\det(J_{g^{-1}}(y,\theta))| \\
    &+ f(x,\theta)|\det(J_{g^{-1}}(y,\theta))|\text{div}(-J_g^{-1}(x,\theta)\partial_\theta g(x,\theta)) \\
    &=|\det(J_{g^{-1}}(y,\theta))|(\diver(-f(x,\theta) J_g^{-1}(x,\theta) \partial_\theta g(x,\theta))+\partial_\theta f(x,\theta)),
\end{align*}
where the last equation follows from the fact that for any real-valued function $h$ and vector-valued function $\vec v$,
\begin{align}\label{eq:diver}
    \diver(h(x)\vec v(x))=\nabla_x h(x)^T \vec v(x) + h(x)\diver( \vec v(x)).
\end{align}

\textbf{\Cref{eq:cov2}:} By \Cref{eq:diver}, we obtain
\begin{align}\label{eq:divy}
\begin{split}
    &~~~~\diver \left(\varphi_n(y) f(g^{-1}(y,\theta),\theta) |\det(J_{g^{-1}}(y,\theta))| \vec v(y)\right)\\
    &= \nabla_y (\varphi_n(y) f(g^{-1}(y,\theta),\theta))^T|\det(J_{g^{-1}}(y,\theta))| \vec v(y) + \varphi_n(y) f(g^{-1}(y,\theta),\theta) \diver(|\det(J_{g^{-1}}(y,\theta))| \vec v(y)).    
\end{split}
\end{align}
Applying \Cref{eq:diver} again, we obtain
\begin{align*}
    \diver(|\det(J_{g^{-1}}(y,\theta))| \vec v(y)) = \nabla_y |\det(J_{g^{-1}}(y,\theta))| ^T \vec v(y) + |\det(J_{g^{-1}}(y,\theta))| \diver(\vec v(y)).
\end{align*}
For $i=1,\cdots,n$,
\begin{align*}
    \partial_{y_i}\det(J_{g^{-1}}(y,\theta))&=\det(J_{g^{-1}}(y,\theta)) \tr(\partial_{y_i} J_{g^{-1}}(y,\theta) J_{g^{-1}}^{-1}(y,\theta))\\
    &=\det(J_{g^{-1}}(y,\theta)) \sum_{j,k}^n \partial_{y_i} (J_{g^{-1}}(y,\theta))_{jk}(J_{g^{-1}}^{-1}(y,\theta))_{kj}\\
    &=\det(J_{g^{-1}}(y,\theta))  \sum_{j=1}^n \sum_{k=1}^n (J_{g^{-1}}^{-1}(y,\theta))_{kj} \partial_{y_k} (J_{g^{-1}}(y,\theta))_{ji}.
\end{align*}
Note that for any differentiable function $h:\R^n\mapsto\R$, $\nabla_x h(g(x,\theta))=J_g(x,\theta)^T\nabla_y h(y)|_{y=g(x,\theta)}$, i.e.,
\begin{align*}
    \nabla_y h(y)|_{y=g(x,\theta)} = J_g^{-1}(x,\theta)^T \nabla_x h(g(x,\theta)).
\end{align*}
Therefore, we can write
\begin{align*}
    \nabla_x (J_g^{-1}(x,\theta))_{ji} = J^{-1}_{g^{-1}}(y,\theta)^T \nabla_y  (J_g^{-1}(g^{-1}(y,\theta),\theta))_{ji},
\end{align*}
i.e., $\partial_{x_j} (J_g^{-1}(x,\theta))_{ji} = \sum_{k=1}^n (J_{g^{-1}}^{-1}(y,\theta))_{kj} \partial_{y_k} (J_{g^{-1}}(y,\theta))_{ji},$ and it follows that
\begin{align*}
    \partial_{y_i}\det(J_{g^{-1}}(y,\theta))&=\det(J_{g^{-1}}(y,\theta)) \sum_{j=1}^n \partial_{x_j} (J_g^{-1}(x,\theta))_{ji}.
\end{align*}
Hence,
\begin{align*}
    &~~~~\nabla_y |\det(J_{g^{-1}}(y,\theta))| ^T \vec v(y)\\
    &=|\det(J_{g^{-1}}(y,\theta))| \sum_{i=1}^n (\sum_{j=1}^n \partial_{x_j} (J_g^{-1}(x,\theta))_{ji}) \vec v_i (y)=\det(J_{g^{-1}}(y,\theta)) \diver (J^{-1}_g(x,\theta)^T)^T \vec v(y),
\end{align*}
where $\diver A(x) := (\sum_{j=1}^n \partial_{x_j} A(x)_{1j},\cdots,\sum_{j=1}^n \partial_{x_j} A(x)_{nj})^T$ for any matrix-valued function $A(x)$. Notice that
%$\diver(J^{-1}_g(x,\theta)^T) = (\sum_{j=1}^n \partial_{x_j} (J_g^{-1}(x,\theta))_{j1},\cdots,\sum_{j=1}^n \partial_{x_j} (J_g^{-1}(x,\theta))_{jn})^T$.
\begin{align*}
    \diver( \vec v(y)) = \tr(\nabla_y \partial_\theta g (g^{-1}(y,\theta),\theta))=\tr(J^{-1}_g(x,\theta)^T\nabla_x \partial_\theta g (x,\theta)),
\end{align*}
where $\nabla_y v(y) := (\nabla_y v_1(y),\cdots,\nabla_y v_n(y))^T$ for any vector-valued function $v(x)$. Therefore,
%$\nabla_y \partial_\theta g (g^{-1}(y,\theta),\theta) = (\nabla_y \partial_\theta g_1 (g^{-1}(y,\theta),\theta),\cdots,\nabla_y \partial_\theta g_n (g^{-1}(y,\theta),\theta))^T$.
\begin{align}
    &~~~~\diver(|\det(J_{g^{-1}}(y,\theta))| \vec v(y))\nonumber\\
    &= |\det(J_{g^{-1}}(y,\theta))| \diver (J^{-1}_g(x,\theta)^T)^T \vec v(y)
    +  |\det(J_{g^{-1}}(y,\theta))|\tr(\nabla_x \partial_\theta g (x,\theta)J^{-1}_g(x,\theta))\nonumber\\
    &= |\det(J_{g^{-1}}(y,\theta))| \diver (J^{-1}_g(x,\theta)^T)^T \vec v(y)
    +  |\det(J_{g^{-1}}(y,\theta))|\tr(J^{-1}_g(x,\theta)\nabla_x \partial_\theta g (x,\theta))\nonumber\\
    &=|\det(J_{g^{-1}}(y,\theta))| \diver(J^{-1}_g(x,\theta) \partial_\theta g (x,\theta))\label{eq:divy1},
\end{align}
where the last equation follows from the divergence formula for matrix-vector production
\begin{align}\label{eq:diver1}
    \diver(A(x)v(x))=\diver(A(x)^T)^T v(x)+\tr(A(x)\nabla v(x)).
\end{align}
Using the chain rule for gradient $\nabla (pq)=p\nabla q+ q \nabla p$, we can write
\begin{align}
    \nabla_y (\varphi_n(y) f(g^{-1}(y,\theta),\theta) ) &=\varphi_n(y) \nabla_y f(g^{-1}(y,\theta),\theta) + f(g^{-1}(y,\theta),\theta)\nabla_y\varphi_n(y)\nonumber\\
    & = J_g^{-1}(x,\theta)^T \nabla_x(\varphi_n(g(x,\theta))f(x,\theta) )\label{eq:divy2}.
\end{align}
Substituting \Cref{eq:divy1,eq:divy2} into \Cref{eq:divy}, we obtain \Cref{eq:cov2}. \hspace*{\fill} $\square$

%\begin{align*}
%    &~~~~\diver (\varphi_n(y) f(g^{-1}(y,\theta),\theta) |\det(J_{g^{-1}}(y,\theta))| \vec v(y))\\
%    &= |\det(J_{g^{-1}}(y,\theta))|  \diver (\varphi_n(g(x,\theta))f(x,\theta) J_g^{-1}(x,\theta) \partial_\theta g(x,\theta)) .
%\end{align*}
\subsection*{Proof of \Cref{thm:glr1}}
\hypertarget{sec:app2}{Since} $\Omega\times\Theta$ is compact and both $f,g$ are continuously differentiable, $\sup_{(x,\theta)\in\Omega\times\Theta}|f(x,\theta)|$ and\\ $\sup_{(x,\theta)\in\Omega\times\Theta}|\det J^{-1}_g(x,\theta)|$ are bounded. Therefore,
\begin{align*}
    \lim_{n\rightarrow\infty}|\E(\varphi(g(X,\theta)))-\E(\varphi_n(g(X,\theta)))|&\leq
    \lim_{n\rightarrow\infty}\int_\Omega |\varphi(g(x,\theta))-\varphi_n(g(x,\theta))| |f(x,\theta)|dx\\
    &\leq\lim_{n\rightarrow\infty}\int_\Omega |\varphi(g(x,\theta))-\varphi_n(g(x,\theta))|dx \sup_{(x,\theta)\in\Omega\times\Theta}|f(x,\theta)|\\
    &\leq\lim_{n\rightarrow\infty}\int_{g(\Omega,\theta)} |\varphi(y)-\varphi_n(y)|dy \sup_{(x,\theta)\in\Omega\times\Theta}|f(x,\theta)\det J^{-1}_g(x,\theta)|=0,
\end{align*}
where the last equation follows from the fact that $\varphi_n\rightarrow\varphi$ in $L^1$.
Similarly, $\sup_{(x,\theta)\in\Omega\times\Theta}|d(x,\theta)+l(x,\theta)|$ is bounded, and
\vspace*{-0.3em}
\begin{align*}
    &~~~~\lim_{n\rightarrow\infty}\left|\int_\Omega \varphi(g(x,\theta)) (d(x,\theta)+l(x,\theta)) f(x,\theta)dx-\int_\Omega \varphi_n(g(x,\theta)) (d(x,\theta)+l(x,\theta)) f(x,\theta)dx\right|\\
    &\leq  \lim_{n\rightarrow\infty}\int_{g(\Omega,\theta)} |\varphi(y)-\varphi_n(y)| dy\sup_{(x,\theta)\in\Omega\times\Theta}|(d(x,\theta)+l(x,\theta))f(x,\theta)\det J^{-1}_g(x,\theta)|=0.
\end{align*}
By Theorem 4 in Section 16.3.5 of \citeN{zorich2004mathematical}, we obtain
\begin{align*}
    \frac{d}{d\theta}\E(\varphi(g(X,\theta)))&=\lim_{n\rightarrow\infty}\frac{d}{d\theta}\E(\varphi_n(g(X,\theta)))\\
    &=\int_\Omega \varphi(g(x,\theta)) (d(x,\theta)+l(x,\theta)) f(x,\theta)dx+\int_{\partial\Omega}\varphi(g(x,\theta))s(x,\theta)^T \vec n(x)f(x,\theta) ds. \tag*{$\square$}
\end{align*}
% Reducing font size (to 9pt) for References & Author Biagraphies
\footnotesize

% Please don't exchange the bibliographystyle style
\bibliographystyle{wsc}

% AUTHOR: Include your bib file here
\bibliography{demobib}

\section*{AUTHOR BIOGRAPHIES}

\noindent {\bf XINGYU REN} is a Ph.D. student in the Department of Electrical and Computer Engineering at the University of Maryland, College Park. His research interests include stochastic optimization and Markov decision processes. His e-mail address is \email{renxy@umd.edu}.\\

\noindent {\bf MICHAEL C. FU} holds the Smith Chair of Management Science in the Robert H. Smith School of Business, with a joint appointment in the Institute for Systems Research and an affiliate appointment in the Department of Electrical and Computer Engineering, at the University of Maryland, College Park. His research interests include stochastic gradient estimation, simulation optimization, and applied probability. He served as WSC2011 Program Chair and received the INFORMS Simulation Society's Distinguished Service Award in 2018. He is a Fellow of INFORMS and IEEE. His e-mail address is \email{mfu@umd.edu}.\\
\end{document}